\documentclass[letterpaper, 10 pt, conference]{ieeeconf}
\IEEEoverridecommandlockouts

\usepackage{amsthm}
\usepackage{cite}
\usepackage{amsmath}
\usepackage{mathtools}
\usepackage{xcolor}
\usepackage{verbatim}
\usepackage[pdftex]{graphicx}
\usepackage{subcaption}
\usepackage{amsfonts}
\usepackage{algorithm}
\usepackage{algpseudocode} 
\usepackage{array}

\DeclareMathOperator{\Diag}{Diag}
\DeclareMathOperator{\tr}{tr}
\DeclareMathOperator{\colspace}{colspace}

\newtheorem{theorem}{Theorem}

\begin{document}

\title{\LARGE \bf Resilient State Estimation in Presence of Severe Coordinated Cyber-Attacks on Large-Scale Power Systems}

\author{Ana Jevti\'c and Marija Ili\'c%
	\thanks{*This material is based upon work supported by the Department of Energy under Award Number DE-OE0000779.}%
	\thanks{Ana Jevti\'c ({\tt\footnotesize ajevtic@mit.edu}) and Marija Ili\'c ({\tt\footnotesize ilic@mit.edu}) are with the Laboratory for Information and Decision Systems (LIDS) at Massachusetts Institute of Technology, 
		Cambridge, MA 02139 USA}%
}

\maketitle

\begin{abstract}
	Providing situational awareness in light of severe coordinated cyber-attacks on power grids, where many measurements may be untrusted, is necessary for reliable monitoring and resilient operation of the grid. In this scenario, the set of good measurements is by itself insufficient for state estimation due to loss of observability. In this paper, we present a resilient state estimation algorithm, based on output clustering. By augmenting the measurement set by respective cluster variables, the system observability is regained, and a reliable state estimate can be computed. We show the numerical performance of our proposed algorithm and its ability to successfully replace corrupted measurements using cluster variables through an example on the IEEE 24-bus power system.
\end{abstract}

\section{Introduction}
The electric power grid has been increasingly exposed to the public via smart devices with Internet connectivity and operation of the grid through corporate networks and use of more open control architectures. However, the evolution of the power grid into a cyber-physical system (CPS) has not been conducted with security in mind.
In the recent years, a number of cyber-incidents have demonstrated the vulnerability of CPS, including the well-known attack on the Ukrainian power grid in 2015 \cite{Ukraine2015}, and the cyber incident that disrupted grid operation in the western US in March of 2019 \cite{westUS}. These events have led to the increase in awareness of the problem of securing critical infrastructure, such as the power grid, transportation systems, gas and water networks, etc. The control systems behind these critical infrastructures have long been protected by physically isolating the local control and communication networks from insecure global networks such as the Internet. Since this physical separation is slowly diminishing, and in the light of these new threats, many attack detection schemes have been proposed in recent years, both in the cyber (intrusion detection systems - IDSs)~\cite{liao2013intrusion,pan2015developing,valenzuela2012real} and the physical layers~\cite{mo2009secure,teixeira2012revealing,jevtic2018physics}.

Further, the problem of state estimation (SE) in presence of cyber-attacks has attracted a lot of attention, since a state estimate is crucial to continued operation of the critical infrastructures. For example, in power grids, the inability to produce a state estimate would cause the Energy Management System (EMS) to be suspended. The resilient SE problem has been formulated as robust SE, both with noiseless \cite{Fawzi2014,Tabuada2016} and noisy measurements \cite{shoukry2015sound}, \cite{pajic2014robustness,farahmand2011doubly}. In \cite{zonouz2012scpse}, the authors propose a fusion framework that leverages the intrusion detection from the cyber-layer, to exclude the compromised measurements from SE, thus producing a reliable estimate. However, these methods are limited by the observability condition. In other words, in the scenario of a large-scale coordinated attack, when many measurements may become unavailable, the system will become unobservable, and it will be impossible to produce a state estimate.

In this paper, we focus on the scenario of severe coordinated cyber-attacks on measurements in power systems. After detecting and localizing the cyber-attacks, too many measurements are excluded from SE, and the system is no longer observable. We propose a clustering-based method for Resilient State Estimation, where the measurement set is augmented, so that system observability is restored. We define this method as resilient since it enables the system operator to maintain situational awareness, and an acceptable level of operational normalcy in response to disturbances including threats of an unexpected and malicious nature \cite{rieger2009resilient}.

The remainder of this paper is structured as follows. Section~\ref{sec:problem} contains the mathematical formulation of the problem considered in this paper. In Section~\ref{sec:RSE}, we present the main contribution of this paper, the Resilient State Estimation method based on output clustering. Section~\ref{sec:simulations} introduces relevant component models and demonstrates the efficiency of our proposed method through a numerical example on the IEEE RTS 24-bus power system. Finally, in Section~\ref{sec:conclusion} we give some concluding remarks.

\textit{Notation}: Let $\mathcal{I}_k$ denote a set of integers, and $|\mathcal{I}_k|$ its cardinality. Then, $e^n_{\mathcal{I}_k} \in \mathbb{R}^{n\times |\mathcal{I}_k|}$ is a matrix composed of column vectors of the identity matrix ${I_n \in \mathbb{R}^{n \times n}}$ corresponding to the index set $\mathcal{I}_k$. We use $\|M\|_F$ to denote the Frobenius norm of matrix $M$. Matrix $M$ is said to be semistable iff the zero eigenvalues of $M$ are semisimple, and all the other eigenvalues have a negative real part. Given a stable proper transfer function of a dynamical system $g(s)$, $\|g(s)\|_{\mathcal{H}_2}$ is the $\mathcal{H}_2$-norm of the system.

\section{Problem Formulation}
\label{sec:problem}
\begin{figure}[h]
	\begin{center}
		\includegraphics[width=0.48\textwidth]{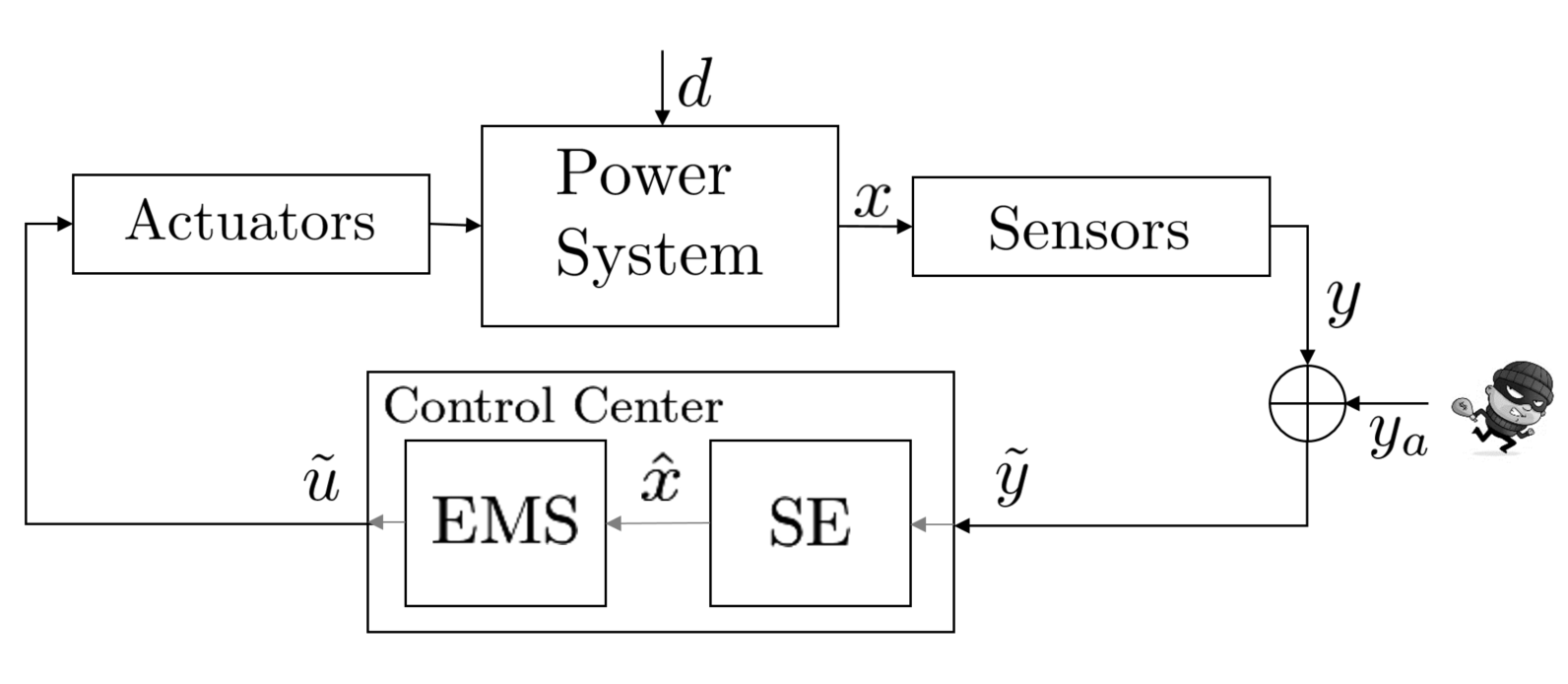}
		\caption{Block diagram of attacked power system. The signal $y_a$ is injected into outputs $y$ to manipulate the system.}
		\label{fig:blkdiag}
	\end{center}
	\vspace{-10pt}
\end{figure}

In this paper, we are concerned with enabling the Control Center to compute a resilient state estimate in presence of severe coordinated cyber-attacks on system measurements. Figure \ref{fig:blkdiag} depicts the block diagram representation of the system we consider. 
\subsection{System Description}
The power system contains various components, such as generators, loads, transmission lines, etc. We assume the physical system follows:
\begin{equation}
	\dot{x}(t)=Ax(t)+B\tilde{u}(t)+d(t)
	\label{eq:phys_dyn}
\end{equation}
where the states of system are denoted $x\in\mathbb{R}^n$, $d(t)$ denotes a disturbance signal, and the control signal issued by the Control Center (CC) is $\tilde{u}\in\mathbb{R}^p$.

 A large network of field sensors is deployed to monitor the operation of the power system in \eqref{eq:phys_dyn}. A malicious attacker can negatively impact the system by manipulating the measurements, which is represented with the added signal $y_a(t)$ in Figure~\ref{fig:blkdiag}:
 \begin{align}
 \tilde{y}(t)&=y(t)+(e^m_A)^Ty_a(t)\\
 y(t)&=Cx(t)
 \label{eq:outputs}
 \end{align}
 where $\tilde{y} \in \mathbb{R}^m$ are measurements received by the CC, and $y_a\in \mathbb{R}^a$ are $a$ attack signals injected by an attacker. Thus, a potentially manipulated measurement signal $\tilde{y}$ reaches the CC, which can then issue a potentially incorrect control signal $\tilde{u}$ to the power system actuators, $ \tilde{u}(t)=K\hat{y}$, where $\hat{y}=C\hat{x}$, and $\hat{x}$ is the state estimate (e.g. from an observer or other state estimation block). Finally, the attacked system can be rewritten in closed-loop as:
\begin{equation}
\Sigma_a:\begin{dcases}
 \dot{x}(t)&=\mathcal{A}x(t)+\mathcal{B}y_a(t)+d(t) \\
\tilde{y}(t)&=Cx(t)+(e^m_A)^Ty_a(t)
\end{dcases}
\label{eq:cl_att_lin_sys}
\end{equation}
where $\mathcal{A}=A+BKC$ is the closed-loop system matrix, and $\mathcal{B}=BK$.
\subsection{Resilient State Estimation Problem}
Under normal conditions (no cyber-attack injected into the system, $\tilde{u}(t)=u(t)$), a linear observer is designed for the system in \eqref{eq:cl_att_lin_sys}, to compute the state estimate $\hat{x}(t)$ from the received measurements:
\begin{equation}
\begin{aligned}
\dot{\hat{x}}(t)&=\mathcal{A}\hat{x}(t)+L(\hat{y}(t)-y(t))\\
&=(\mathcal{A}-LC)\hat{x}(t)+Ly(t)\\
\hat{y}(t)&=C\hat{x}(t) 
\end{aligned}
\label{eq:typ_observer}
\end{equation}
In order to generate a state estimate $\hat{x}$, used for control purposes, it is necessary and sufficient for the pair $(\mathcal{A},C)$ to be observable. This condition holds during normal operation of power systems, even in presence of sensor failures, due to redundancy in measurements. In other words, when such an equipment failure occurs, the affected measurement is simply removed from state estimation, and the state estimate $\hat{x}$ is computed from the remaining (good) measurements. Similar procedure can be applied in the event of a cyber-attack. Assuming that attack detection and localization schemes are in place (such as in \cite{jevtic2018physics,jevtic2020clustering}), both in  cyber and physical layers, the attacked measurements can be removed from state estimation. This implies that $e^m_A$ is known, and that the measurement matrix $C$ can be decomposed as 
${C=\begin{bmatrix}
C_1 & C_A
\end{bmatrix}^T\pi}$
, where $C_1 \in \mathbb{R}^{(m-a)\times n}$ corresponds to trusted measurements, $C_A \in \mathbb{R}^{a\times n}$ to the attacked measurements, and $\pi$ is a permutation matrix. Without loss of generality, we assume that measurements are already ordered in this fashion, i.e. $\pi=I$. A state estimate can then be produced if and only if $(\mathcal{A},C_1)$ is still observable. However, during severe coordinated cyber-attacks, too many measurements may be compromised. Thus, it will not be possible to produce a state estimate.

In this paper, we address the problem of providing a state estimate in the situation of severe coordinated cyber-attacks. We do so by constructing a matrix $\overline{C}_A$, such that, with an augmented matrix ${\overline{C}=\begin{bmatrix}
	C_1 & \overline{C}_A
	\end{bmatrix}^T}$, the pair $(\mathcal{A},\overline{C})$ is observable, and a state estimate can be computed using the augmented set of measurements
\begin{equation}
\bar{y}=\overline{C}x
\label{eq:aug_meas}
\end{equation}
Using this definition, the error dynamics and residual for the system in \eqref{eq:typ_observer} using augmented measurements \eqref{eq:aug_meas} can be written as:
\begin{align}
\dot{e}(t)&=(\mathcal{A}-LC)\hat{x}(t)-(\mathcal{A}-L\overline{C})x(t)-d(t)\\
r(t)&=\hat{y}(t)-\bar{y}(t)
\label{eq:res_obsv}
\end{align}

\begin{figure}[t]
	\centering
	\begin{subfigure}[h]{0.24\textwidth}
		\includegraphics[scale=0.55]{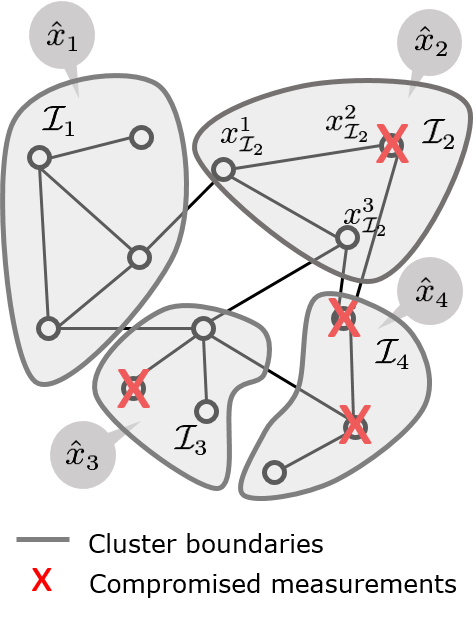}
		\caption{}
		\label{fig:clusters}
	\end{subfigure}~
	\begin{subfigure}[h]{0.24\textwidth}
		\includegraphics[scale=0.55]{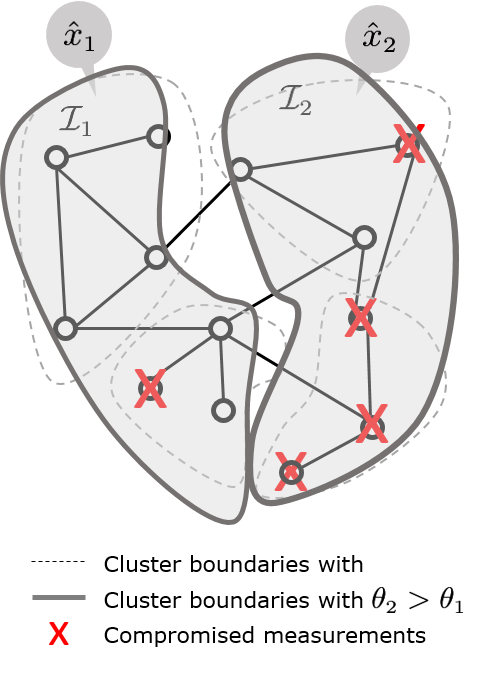}
		\caption{}
		\label{fig:clusters_exp}
	\end{subfigure}
	\caption{\textbf{a)} Example of cluster boundaries; $\hat{x}_2$ is used instead of $x_{\mathcal{I}_2}^2$ to produce a state estimate (similarly for other clusters). \textbf{b)} Example of cluster boundaries for a larger attack; parameter $\theta$ is used to ensure each cluster contains at least one trusted measurement.}
\end{figure}
The rest of this paper will address the construction of matrix $\overline{C}$ based on clustering the system outputs. The aim is to first aggregate measurements with similar dynamic responses into clusters, where the aggregate behavior of each cluster is a close approximation of the measurements within it. Then, instead of each of the attacked measurements, the aggregate behavior of the cluster it belongs to can be used as a surrogate during the state estimation process (depicted in Figure~\ref{fig:clusters}). As the clustering-based aggregation is an approximation procedure, the resulting state estimate will be less accurate than if all the measurements were available, but it is an important trade-off that must be made in order to retain a necessary level of situational awareness during severe cyber incidents. Therefore, this method is not intended to replace the currently used state estimation methods, but to supplement it during critical events.

\section{Clustering-based Resilient State Estimation}
\label{sec:RSE}
In this section, we introduce the clustering procedure on the system $\Sigma$, and the construction of matrix $\overline{C}_A$ based on this clustering.
\subsection{Measurement clustering procedure}
Consider the system $\Sigma$ in normal operation, in absence of cyber-attacks ($y_a(t)\equiv 0$):
\begin{equation}
\Sigma:\begin{dcases}
\dot{x}(t)&=\mathcal{A}x(t)+d(t) \\
y(t)&=Cx(t)
\end{dcases}
\label{eq:cl_lin_sys}
\end{equation}
To quantify the behavior of measurement signals, and the aforementioned similarity between them, we first define clusters $\mathcal{I}_k$ as disjoint subsets of $\mathbb{L}$, where $\mathbb{L}=\{1,\dots, l\}$ is the set of measurement indices. More specifically, clusters are subsets of measurements that have a similar trajectories in time domain. Measurements $i,j$ belonging to the cluster $\mathcal{I}_k$ are approximately proportional $a_iy_i(t) \approx a_jy_j(t) \approx \dots \approx z^{(k)}(t)$, where $a_i, a_j, \dots$ are constant coefficients.
We aim to estimate the full system state based on the combination of received trusted measurements and the hidden system structure contained in the clustered representation of the system. 
With this intuition in mind, we aim to partition the set $\mathbb{L}$ into clusters $\mathcal{I}_k$ such that
\begin{equation}
p_jg_i(s)=p_ig_j(s), \quad \forall i,j \in \mathcal{I}_k
\end{equation}
where $g_i$ is the $i$-th element of $g(s)=C(sI_n-\mathcal{A})^{-1}$, a transfer matrix of the system in \eqref{eq:cl_lin_sys}. We can rewrite the condition for cluster formation more compactly as
\begin{equation}
	(e_{\mathcal{I}_k}^n)^Tg(s)=p_k^T\bar{g}(s)
	\label{eq:equal_gs}
\end{equation} 
where $\bar{g}(s)$ is a scalar function.
This definition provides intuition on the meaning of clustering in our application, but is not practical for designing a procedure that would form such clusters, which would require performing similarity checks on functions. To get around this problem, we will derive a matrix-based condition equivalent to \eqref{eq:equal_gs}, based on the notion of observability. To that end, we first derive the observability Gramian of a semistable system \eqref{eq:cl_lin_sys}. The observability Gramian is defined as \cite{ChenLinSysBook}
\begin{equation}
	W_o=\int_{0}^{\infty}e^{\mathcal{A}^Tt}C^TCe^{\mathcal{A}t}dt
	\label{eq:obsvGram}
\end{equation}
When $\mathcal{A}$ is Hurwitz, the above integral converges, and $W_o$ can also be found as a solution of the Lyapunov equation $\mathcal{A}^TW_o+W_o\mathcal{A}+C^TC=0$.
However, in power systems, the system matrix $\mathcal{A}$ has an inherent structural singularity, as a direct consequence of power conservation law. Due to semistability of the system matrix $\mathcal{A}$, the integral in \eqref{eq:obsvGram} may not converge. Thus, we consider the decomposition of $\mathcal{A}=U\Lambda V^{-1}$, where $U=[u_{max} \quad \bar{U}]$ and $V=[v_{max} \quad \bar{V}]^T$, and $u_{max}$ and $v_{max}$ are the right and left eigenvectors corresponding to the largest eigenvalue ($\lambda_1=0$). Let $\bar{\mathcal{A}}=\overline{V}^T\mathcal{A}\overline{U}$ and $\overline{C}=C\overline{U}$, defined as the stable subspace of $\Sigma$. Then, the observability Gramian of the semistable system is 
\begin{equation}
W_o=\overline{U}\;\overline{W}_o\overline{U}^T
\label{eq:semisGram}
\end{equation}
where $\overline{W}_o$ is the observability Gramian associated with the stable subspace $(\bar{\mathcal{A}},\overline{C})$ of $\Sigma$. 
In the following theorem we find the condition equivalent to \eqref{eq:equal_gs} using the observability Gramian $W_o$ of the semistable system $\Sigma$.
 \begin{theorem}
	Consider the observability Gramian $W_o$ in \eqref{eq:semisGram} of the semistable system $\Sigma$ in \eqref{eq:cl_lin_sys}. Furthermore, let the Cholesky factorization of $W_o$ be given by ${W_o=W_LW_L^T}$, and ${\Phi=W_L}$. Then, the condition in \eqref{eq:equal_gs} is equivalent to 
	\begin{equation}
		(e_{\mathcal{I}_k}^n)^T\Phi=p_k^T\bar{\phi}	
		\label{eq:equal_ws}
	\end{equation}
	where $\bar{\phi}\in\mathbb{R}^{1\times n}$ is a constant vector.
\end{theorem}
\begin{proof}
	See Appendix.
\end{proof}
However, in real systems, the identity in \eqref{eq:equal_gs} is almost never the case. Therefore, we relax the strict equality, and require 
\begin{equation}
	\|p_jg_i(s)-p_ig_j(s)\|_{\mathcal{H}_2} \leq \varepsilon, \quad \forall i,j \in \mathcal{I}_k
	\label{eq:clustering_condition}
\end{equation}
to hold for each cluster. Equivalently, we can check for linear dependence between rows of matrix $\Phi$:
\begin{equation}
\|p_j\Phi_i-p_i\Phi_j\| \leq \theta \quad \forall i,j \in \mathcal{I}_k
\label{eq:clust_cond}
\end{equation}
where $\theta >0$ and $\Phi_i$ is the $i$-th row of $\Phi$. Here, $\theta$ is a parameter that allows us to control the coarseness of clustering. In other words, it allows us to find outputs that have a "similar", instead of equal, response, which relaxes the condition \eqref{eq:equal_gs}. However, the choice of $\theta$ is not trivial, as it introduces a trade-off between accuracy of the approximation and size of clusters. In general, $\theta$ should be chosen as a smallest value for which each cluster contains at least one trusted measurement (as depicted in Figure~\ref{fig:clusters_exp}).
  
\subsection{Construction of matrix $\overline{C}_A$}
After the clusters have been defined, we can construct the matrix $\overline{C}_A$ that will be used to augment the set of available trusted measurements so that the system is observable. Then, the system operator can be provided with situational awareness using the resilient state estimate.
In the analysis in previous section, we have shown that clusters can be formed such that measurements $i,j$ within the cluster $\mathcal{I}_k$ are approximately proportional, i.e. $a_iy_i(t) \approx a_jy_j(t) \approx \dots \approx z^{(k)}(t)$. Then, we derived a matrix-based condition to find such clusters. Next, we show that the 
state estimate can be produced using the augmented matrix $\overline{C}$, by choosing $\overline{C}_A=(e_A^m)^T\Pi^T\Pi C$. The clustering matrix $\Pi \in \mathbb{R}^{K\times n}$ is defined as:
\begin{equation}
\Pi:=\Diag\{p_1,p_2,\dots, p_K\}E \in \mathbb{R}^{K \times n}
\label{eq:agg_matrix}
\end{equation}
where $E$ is a permutation matrix and $p_k$ are clustering coefficients. 
The residual $r=\hat{y}-\bar{y}$ defined in \eqref{eq:res_obsv} will converge to 0 if the error system $g_{\hat{y}}-g_{\bar{y}}$ also converges to 0. The transfer matrix associated with $\hat{y}$ is $g_{\hat{y}}(s)=g_y(s)=C(sI-\mathcal{A})^{-1}$, and the transfer matrix associated with $\bar{y}$ is ${g_{\bar{y}}(s)=\Pi^T\Pi C\left(sI_n-\mathcal{A}\right)^{-1}}$. The following theorem establishes the convergence of the error system.
\begin{theorem}
	Consider a semistable linear system in \eqref{eq:cl_lin_sys} and the augmented set of measurements $\bar{y}$ in \eqref{eq:aug_meas}.  Then, the error system $g_e(s)=g_{\hat{y}}(s)-g_{\bar{y}}(s)$ is asymptotically stable, and state $x$ can be estimated using measurements $\bar{y}$.
\end{theorem}
\begin{proof}
	See Appendix.
\end{proof}

\section{Numerical examples}
\label{sec:simulations}
We begin this section by providing necessary power system component models, and deriving the standard state space model of the interconnected power system in form given in \eqref{eq:cl_lin_sys}. Then, we demonstrate the performance of our method on the IEEE RTS 24-bus system.
\subsection{Power system modeling}
\label{sec:modeling}
In this section, we introduce the power system component models used to derive the system matrices in \eqref{eq:cl_lin_sys}. We model the loads as dynamic using the structure-preserving load model, alongside the generator model with governor control.
\par We consider a power system with $n_G$ generators and $n_L$ loads, and denote the set of generator buses by $\mathcal{G}$, and the set of load buses by ${\mathcal{L}}$. The mechanical dynamics of generators with governor control and aggregate loads at the substation level are given by:
\begin{equation}
\begin{aligned}
\label{eq:mech_dyn}
J_i\dot{\omega}_i+D_i\omega_i&=P_{T,i}-P_i+e_{T,i}a_i,&& i\in\mathcal{G}\\ 
J_i\dot{\omega}_i+D_i\omega_i&=-P_i-L_i,&& i\in\mathcal{L}\\
T_{u,i}\dot{P}_{T,i}&=-P_{T,i}+K_{t,i}a_i,&&i\in\mathcal{G} \\
T_{g,i}\dot{a}_i&=-r_ia_i-(\omega_i-\omega^{ref}),&&i\in\mathcal{G}
\end{aligned}
\end{equation}
For each bus $i$, state variable $\omega_i$ denotes its frequency, $P_i$ the net real power injected into the network, and states $P_{T,i}$ and $a_i$ denote the mechanical power of the generator and the turbine valve position. At load buses $i \in \mathcal{L}$, $L_i$ is defined as actual mechanical power consumed by the load.  $\omega^{ref}$ is the frequency reference provided by the higher control layer. In order to derive the interconnected system, we treat $P_i$ as a coupling state variable whose dynamics can be obtained by differentiating the linearized DC power flow equation, expressed in matrix form as:
\begin{equation}
\begin{bmatrix} \dot{P}_G \\ -\dot{P}_L\end{bmatrix}=Y_{bus}\mathbf{\omega} \text{  , where  }  Y_{bus}=\begin{bmatrix}Y_{GG} & Y_{GL}\\
Y_{LG} & Y_{LL}
\end{bmatrix} 
\label{eq:net_dyn}
\end{equation} 
where $P_G:=[P_i]_{i\in\mathcal{{G}}}$ and $P_L:=[P_i]_{i\in\mathcal{{L}}}$, and $Y_{bus}$ is the admittance matrix of a lossless transmission network. \par
In the following section, we present numerical simulation examples performed on the IEEE RTS 24-bus power system to illustrate the performance of our proposed resilient SE method. 
\subsection{Test system and illustrative scenarios}
\begin{figure*}[t]
	\centering
	\begin{subfigure}[h]{0.38\textwidth}
		\includegraphics[scale=0.29]{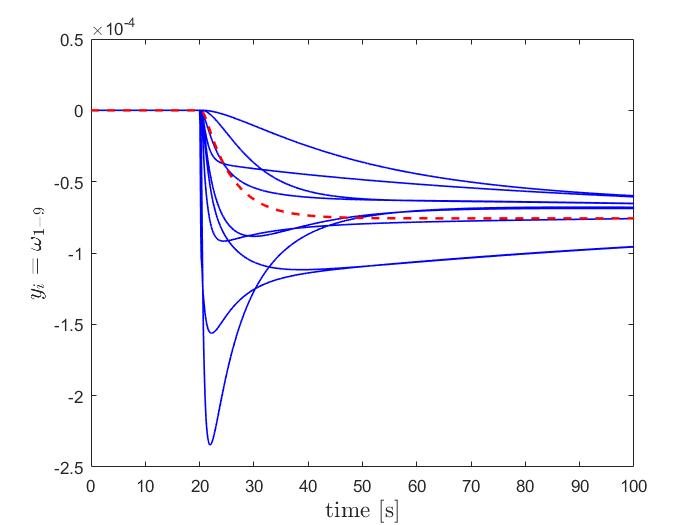}
		\caption{Cluster of frequency measurements}
		\label{fig:freq_clus}
	\end{subfigure}
~
	\begin{subfigure}[h]{0.38\textwidth}
		\includegraphics[scale=0.29]{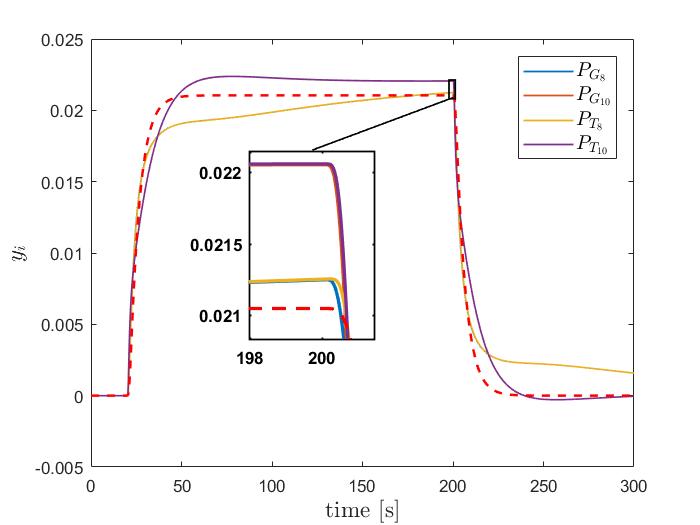}
		\caption{Cluster of power measurements}
		\label{fig:pow_clus}
	\end{subfigure}
	~
	\begin{subfigure}[h]{0.19\textwidth}
		\includegraphics[scale=0.5]{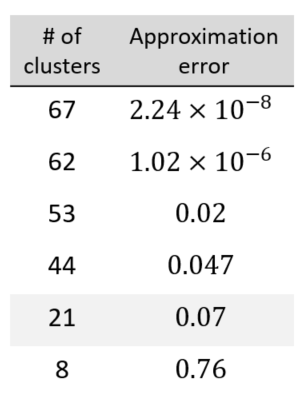}
		\caption{Approximation error for different size clusters}
		\label{fig:table}
	\end{subfigure}
	\caption{Two examples of clusters in the IEEE RTS 24-bus system. In (a) and (b), solid lines are real measurements, red dotted line is the approximated cluster measurement to be used as surrogate in case of a cyber-attack. Table in (c) compares the approximation error depending on the number of clusters that are formed.}
	\label{fig:measurements}
\end{figure*}

The IEEE RTS 24-bus system \cite{IEEE24bus} consists of 10 generators, equipped with governor control, and 14 loads. The interconnected system is modeled using equations \eqref{eq:mech_dyn}-\eqref{eq:net_dyn}, where the dimension of $x$ is 68. In Figure~\ref{fig:measurements}, we analyze the system in the following scenario. From $t=0$ to $20$ s, the loading is nominal. At time $t=20$ s, load at bus 3 increases by $0.1$ p.u., and at time $t=200$ loading returns to nominal value. \par 
Under this scenario, we plot the trajectories of measurements (solid lines) within two clusters, one containing frequency, and one containing power measurements. In Figures \ref{fig:freq_clus} and \ref{fig:pow_clus}, we plot in dotted line the cluster variable, i.e. the surrogate to be used in place of any measurement within the cluster in case of a cyber-attack. In both cases, the cluster variable resembles a centroid of the measurements within it, and can be used for resilient state estimation. In Figure~\ref{fig:table}, we demonstrate the accuracy of state estimation using cluster variables. Specifically, we compare the approximation error introduced by clustering, depending on the number of clusters that are formed. Recall that clusters are disjoint subsets of measurements, i.e. large number of clusters means that the clusters are very small in size, and vice versa. Figures \ref{fig:freq_clus} and \ref{fig:pow_clus} present the scenario where measurements were clustered into 21 clusters, which resulted in approximation error of $\sim 7\%$. This choice allows for larger clusters, containing more than one measurement, while maintaining accuracy of approximation and, therefore, suitable for our proposed resilient state estimation method.

\section{Conclusion}
\label{sec:conclusion}
In this paper, we presented a resilient state estimation algorithm, based on dynamic clustering, which maintains situational awareness during severe coordinated cyber-attacks on system measurements. The set of good measurements, itself insufficient for state estimation, was augmented by respective cluster variables, to provide a reliable state estimate in critical situations. We numerically showed the performance of our proposed algorithm and its ability to successfully replace corrupted measurements using cluster variables through an example on the IEEE 24-bus power system.
\appendix
\begin{proof}[Proof of Theorem 1]
	In order for \eqref{eq:equal_gs} to hold, for each ${i,j \in \mathcal{I}_k}$ it must hold that $p_j\|g_i(s)\|_{\mathcal{H}_2}=p_i\|g_j(s)\|_{\mathcal{H}_2}.$ Similarly, \eqref{eq:equal_ws} is equivalent to $p_j\|\Phi_i\|=p_i\|\Phi_j\|$, where $\Phi_i$ is the $i$th row of the matrix $\Phi$. The $\mathcal{H}_2$-norm of a linear system can be computed as the $\mathcal{L}_2$-norm of its impulse response $h(t)$.
	$$\|g(s)\|_{\mathcal{H}_2}^2=\|h(t)\|_{2}^2=\tr\left\{\overline{U}\left[\int_{0}^{\infty}e^{\bar{\mathcal{A}}^Tt}\overline{C}^T\overline{C}e^{\bar{\mathcal{A}}t}dt\right]\overline{U}^T\right\}$$ 
	For $\|h(t)\|_{2}^2$ to be finite, the integral above must be finite. Since $\bar{\mathcal{A}}$ and $\overline{C}$ are the stable subspace of $\Sigma$, we have
	$\lim\limits_{t\rightarrow\infty}e^{\bar{\mathcal{A}}t}=0$. Therefore,  $\|h(t)\|_{2}^2$ is finite and equal to:
	\begin{align*}
	\|g(s)\|_{\mathcal{H}_2}^2&=\|h(t)\|_{2}^2=\tr\{W_o\}=\tr\{W_LW_L^T\}= \nonumber \\ 
	&=\|W_L\|_F=\|\Phi\|_F
	\end{align*}	
	where $\|\cdot\|_F$ is a vector norm applied to each row of $\Phi$.
	Hence, \eqref{eq:equal_gs} is equivalent to \eqref{eq:equal_ws}.
\end{proof}
\begin{proof}[Proof of Theorem 2]
	By definition, $\Pi$ is a unitary matrix. Also, by definition, $v_{max} \in \colspace(\Pi^T)$. Let $\bar{\Pi}$ be an orthogonal complement of $\Pi$. Therefore, $I-\Pi^T\Pi=\bar{\Pi}^T\bar{\Pi}$.
	Consider now the error system $g_e$:
	\begin{equation}
	\begin{aligned}
	g_e(s)&=C(sI-\mathcal{A})^{-1}-\Pi^T\Pi C(sI-\mathcal{A})^{-1}= \\
	&=(I_n-\Pi^T\Pi)C(sI-\mathcal{A})^{-1}=\bar{\Pi}^T\bar{\Pi}g(s)
	\end{aligned}
	\end{equation}
	We have $\Pi^T\Pi v_{max}=v_{max}$, or equivalently $\bar{\Pi}v_{max}=0.$
	This implies that there is pole-zero cancellation in $\bar{\Pi}g(s)$ associated with the zero eigenvalue. Therefore, all poles of $\bar{\Pi}g(s)$ have negative real parts, and the error system $g_e$ is asymptotically stable.
\end{proof}

\bibliographystyle{unsrt}
\bibliography{PESGM2020}
\end{document}